\documentclass{emulateapj}
\usepackage{apjfonts}

\shorttitle{The Mass--Metallicity Relation}
\shortauthors{Tremonti et al.}

\newcommand{\epssz}{\epsscale{1.2}}

\begin{document}

\title{The Origin of the Mass--Metallicity Relation: \\
Insights from 53,000 Star-Forming Galaxies in the SDSS}

\author{Christy A.\ Tremonti\altaffilmark{1,2}, 
Timothy M.\ Heckman\altaffilmark{1},
Guinevere Kauffmann\altaffilmark{3}, 
Jarle Brinchmann\altaffilmark{3,4}, 
St\'{e}phane Charlot\altaffilmark{3,5},
Simon D.~M.\ White\altaffilmark{3}, 
Mark Seibert\altaffilmark{1,6},
Eric W.\ Peng\altaffilmark{1,7}, 
David J.\ Schlegel\altaffilmark{8},
Alan Uomoto\altaffilmark{1,9}
Masataka Fukugita\altaffilmark{10}, and
Jon Brinkmann\altaffilmark{11}
}

\altaffiltext{1}{Department of Physics and Astronomy, 
  The Johns Hopkins University, 3400 N.\ Charles Street, Baltimore, MD
  21218, USA}
\altaffiltext{2}{Steward Observatory, 933 N.\ Cherry Ave., Tucson, AZ
  85721, USA}
\altaffiltext{3}{Max Planck Institut f\"{u}r Astrophysik, 
  Karl-Schwarzschild-Str. 1 Postfach 1317, D-85741 Garching, Germany}
\altaffiltext{4}{Centro de Astrof{\'\i}sica da Universidade do Porto, 
Rua das Estrelas - 4150-762 Porto, Portugal}
\altaffiltext{5}{Institut d'Astrophysique de Paris, CNRS, 
  98 bis Boulevard Arago, 75914 Paris, France}
\altaffiltext{6}{Department of Astronomy, California Institute of
Technology, 105-24 Caltech, 1201 East California Blvd, Pasadena, CA
  91125, USA}
\altaffiltext{7}{Department of Physics and Astronomy, Rutgers, 
the State University of New Jersey, 136 Frelinghuysen Road,
Piscataway, NJ 08854-8019, USA}
\altaffiltext{8}{Princeton University Observatory, Peyton Hall,
  Princeton, NJ 08544-1001, USA}
\altaffiltext{9}{Carnegie Observatories, 813 Santa Barbara Street
Pasadena, California 91101, USA}
\altaffiltext{10}{Institute for Cosmic Ray Research, University of
  Tokyo, Kashiwa 277-8582, Japan}
\altaffiltext{11}{Apache Point Observatory, PO Box 59, Sunspot, NM
  88349, USA}

\begin{abstract}

We utilize Sloan Digital Sky Survey imaging and spectroscopy of
$\sim$53,000 star-forming galaxies at \mbox{$z\sim0.1$} to study the
relation between stellar mass and gas-phase metallicity.  We derive
gas-phase oxygen abundances and stellar masses using new techniques
which make use of the latest stellar evolutionary synthesis and
photoionization models.  We find a tight ($\pm$0.1~dex) correlation
between stellar mass and metallicity spanning over 3 orders of
magnitude in stellar mass and a factor of 10 in metallicity. The
relation is relatively steep from $10^{8.5}$ -
$10^{10.5}$~M$_{\sun}h_{70}^{-2}$, in good accord with known trends
between luminosity and metallicity, but flattens above
$10^{10.5}$~M$_{\sun}$.  We use indirect estimates of the gas mass
based on the H$\alpha$ luminosity to compare our data to predictions
from simple closed box chemical evolution models.  We show that metal
loss is strongly anti-correlated with baryonic mass, with low mass
dwarf galaxies being 5 times more metal-depleted than L$^{*}$ galaxies
at $z\sim0.1$.  Evidence for metal depletion is not confined to dwarf
galaxies, but is found in galaxies with masses as high as
$10^{10}$~M$_{\sun}$.  We interpret this as strong evidence both of
the ubiquity of galactic winds and of their effectiveness in removing
metals from galaxy potential wells.

\end{abstract}


\section{Introduction:}

Stellar mass and metallicity are two of the most fundamental physical
properties of galaxies.  Both are metrics of the galaxy evolution
process, the former reflecting the amount of gas locked up into stars,
and the latter reflecting the gas reprocessed by stars and any
exchange of gas between the galaxy and its environment.  Understanding
how these quantities evolve with time and in relation to one another
is central to understanding the physical processes that govern the
efficiency and timing of star formation in galaxies.

The influence of stellar winds and supernovae on the interstellar
medium (ISM) of galaxies, generally dubbed `feedback', has been
regarded as an important ingredient in galaxy evolution since the
1970s \citep{Larson_1974, Larson_and_Dinerstein_1975,
White_and_Rees_1978}.  Feedback is believed to play a critical role in
regulating star formation by reheating the cold ISM and by physically
removing gas from the disk and possibly the halo via galactic winds.
Various lines of observational evidence have established that
large-scale outflows of gas are ubiquitous among the most actively
star-forming galaxies at low and high redshift
\citep[e.g.][]{Lehnert_and_Heckman_1996, Dahlem_et_al_1998,
Rupke_et_al_2002, Shapley_et_al_2001, Frye_et_al_2002}.  However,
detailed studies of winds in nearby starbursts have shown them to be a
complex, multiphase, hydrodynamical phenomenon with the majority of
the energy and newly synthesized metals existing in the
hard-to-observe coronal and hot phases (T = $10^5 - 10^7$~K)
\citep[e.g.][]{Strickland_et_al_2002}.  This complexity has prevented
both a direct assessment of the cosmological impact of galactic winds,
and the development of physically accurate prescriptions for
incorporating feedback into semi-analytical and numerical models of
galaxy formation.

Fortunately it is possible to obtain some quantitative information
about the \emph{impact} of galactic winds without a full understanding
of the abstruse physics responsible for their morphology and
kinematics.  Galaxies which host winds powerful enough to overcome the
gravitational binding energy of their halos will vent some of their
metals into the intergalactic medium.  Hence, one way of evaluating
the importance of galactic winds is to look for their chemical imprint
on galaxies.  However, low metallicity is not necessarily a hallmark
of wind activity.  The metallicity of a galaxy is expected to depend
strongly on its evolutionary state, namely how much of its gas has
been turned into stars.  To detect metal \emph{depletion} it is
therefore necessary to make some assumptions about the expected level
of chemical enrichment based on a galaxy's star and gas content.  Of
course other mechanisms besides winds could make a galaxy appear to be
metal-depleted --- for example, the inflow of pristine gas, or the
return of comparatively unenriched material from evolved low mass
stars.  But these scenarios can potentially be distinguished by
examining the dependence of metal depletion on dynamical mass.  To
quantify the impact of feedback on the local galaxy population we
therefore compare the observed metallicities of galaxies spanning a
wide range in total mass to the predictions of simple chemical
evolution models.

Interest in the relationship between mass and metallicity dates back
several decades, beginning with the seminal work of
\citet{Lequeux_et_al_1979}.  However, because of the difficulty of
obtaining masses, luminosity was generally adopted as a surrogate.  A
correlation between metallicity and blue luminosity was demonstrated
by \citet{Garnett_and_Shields_1987} and extended by various authors
\citep[e.g][]{Skillman_et_al_1989, Brodie_and_Huchra_1991,
Zaritsky_Kennicutt_and_Huchra_1994} to include a range of Hubble types
and to span over 11 magnitudes in luminosity and 2~dex in metallicity.
The interpretation of this striking trend has, however, remained a
matter of some debate owing to the difficulty of transforming from
observables to the ingredients needed for simple chemical evolution
models (stellar mass, gas mass, and metallicity).

In recent years, the development of more sophisticated models for
stellar populations \citep[e.g.][]{Bruzual_and_Charlot_2003} and
gaseous nebulae \citep[e.g.][]{Ferland_1996,
Charlot_and_Longhetti_2001} has resulted in major advances in our
ability to derive physical properties from observables. In this work
we derive stellar mass-to-light (M/L) ratios according to the methods
of \citet{Kauffmann_et_al_2003a} which rely on spectroscopic line
indices to help circumvent the classical age-metallicity-reddening
degeneracy issues.  We measure metallicity using the formalism of
\citet{Charlot_et_al_2004} which makes use of all of the strong
optical nebular lines in our bandpass (as opposed to more traditional
methods which rely on a single line ratio.)  We couple these improved
techniques with the enormous statistical power provided by the Sloan
Digital Sky Survey (SDSS). We present the mass--metallicity relation of
$\sim53,000$ star-forming galaxies at $z\sim0.1$ as a new benchmark
for successful models of galaxy evolution.

We begin with a brief description of the SDSS data products and our
data processing techniques in \S\ref{data}.  We outline our method for
measuring metallicity and stellar mass in \S\ref{measurements}.  In
\S\ref{lummetal} we compare the luminosity--metallicity relation of the
SDSS data with previous determinations.  We present the
mass--metallicity relationship in \S\ref{massmetal} and consider its
origin in \S\ref{mzorigin}.  We explore sources of systematic error in
\S\ref{error} and conclude in \S\ref{discussion}.  Where appropriate
we adopt a cosmology of $\Omega_{\textrm{M}} = 0.3$, 
$\Omega_{\Lambda} = 0.7$, and H$_0$ = 70 km~s$^{-1}$ Mpc$^{-1}$.

\section{The SDSS Data}\label{data}

The data analyzed in this study are drawn from the Sloan Digital Sky
Survey (SDSS).  The survey goals are to obtain photometry of a quarter
of the sky and spectra of nearly one million objects.  Imaging is
obtained in the \emph{u, g, r, i, z} bands \citep{Fukugita_et_al_1996,
Smith_et_al_2002} with a special purpose drift scan camera
\citep{Gunn_et_al_1998} mounted on the SDSS 2.5~meter telescope at
Apache Point Observatory.  The imaging data are photometrically
\citep{Hogg_et_al_2001} and astrometrically \citep{Pier_et_al_2003}
calibrated, and used to select stars, galaxies, and quasars for
follow-up fiber spectroscopy.  Spectroscopic fibers are assigned to
objects on the sky using an efficient tiling algorithm designed to
optimize completeness \citep{Blanton_et_al_2003a}.  The details of the
survey strategy can be found in \citet{York_et_al_2000} and an
overview of the data pipelines and products is provided in the Early
Data Release paper \citep{Stoughton_et_al_2002}.

Our parent sample for this study is composed of 211,265 objects which
have been spectroscopically confirmed as galaxies and have data
publicly available in the SDSS Data Release~2
\citep[DR2;][]{Abazajian_et_al_2004}.  These galaxies are part of the
SDSS `main' galaxy sample used for large scale structure studies
\citep{Strauss_et_al_2002} and have Petrosian $r$ magnitudes in the
range $14.5 < r < 17.77$ after correction for foreground galactic
extinction using the reddening maps of
\citet{Schlegel_Finkbeiner_and_Davis_1998}.  Their redshift
distribution extends from $\sim0.005$ to 0.30, with a median $z$ of 0.10.
From this sample we select a subset of 53,400 star-forming galaxies
for nebular analysis, as discussed in \S\ref{selection}.

We determine stellar mass-to-light ratios and nebular metallicities
for our sample galaxies from the SDSS spectra. The spectra are
obtained with two 320-fiber spectrographs mounted on the SDSS 
2.5-meter telescope.  Fibers 3\arcsec\ in diameter are manually plugged
into custom-drilled aluminum plates mounted at the focal plane of the
telescope. The spectra are exposed for 45 minutes or until a fiducial
signal-to-noise (S/N) is reached.  The median S/N per pixel for
galaxies in the main sample is $\sim14$.  The spectra are processed by
an automated pipeline (Schlegel et al., in prep.)  which flux and
wavelength calibrates the data from 3800 to 9200~\AA.  The
instrumental resolution is R~$\equiv \lambda/\delta\lambda$ = 1850 --
2200 (FWHM$\sim2.4$~\AA\ at 5000~\AA).

The wavelength coverage, resolution, S/N, and general high quality of
the SDSS spectra make them extremely well suited for the derivation of
nebular abundances.  However the spectrophotometric calibration and
the small size of the fiber aperture relative to the target galaxies
pose some concern.  The SDSS spectrographs do not employ an
atmospheric dispersion corrector and the spectra are frequently
acquired under non-photometric conditions.  The Survey has
nevertheless been able to obtain a remarkable level of
spectrophotometric precision by the simple practice of observing
multiple standard stars simultaneously with the science targets.  (The
artifice in this case is that the `standards' are not classical
spectrophotometric standards, but are halo F-subdwarfs that are
calibrated to stellar models -- see \citet{Abazajian_et_al_2004} for
details.)  To quantify the quality of the spectrophotometry we have
compared magnitudes synthesized from the spectra with SDSS photometry
obtained with an aperture matched to the fiber size. The 1$\sigma$
error in the synthetic colors is 5\% in $g-r$ and 3\% in $r-i$
($\lambda_{g}\sim4700$~\AA; $\lambda_{r}\sim6200$~\AA;
$\lambda_{i}\sim7500~$\AA).  At the bluest wavelengths
($\sim3800$~\AA) we estimate the error to be $\sim12$\% based on
repeat observations.  There is also a systematic error in the sense
that the spectra are bluer than the imaging by $\sim$2\% in the
$g$-band, but it is unclear at present whether this represents an
error in the absolute calibration of the photometry or the
spectroscopy.  While the random errors in the spectrophotometry are
offset by the large size of the data set, the accuracy of our measured
nebular abundances relative to true global abundances is compromised
to some degree by the size of the fiber aperture. At the median
redshift of the SDSS main galaxy sample ($z\sim0.1$) the target
galaxies subtend several arcseconds on the sky.  The fraction of
galaxy light falling in the 3\arcsec\ fiber aperture is typically
about 1/3, with variations in redshift and morphological type causing
the fraction to range from 0.05 -- 0.6.  The impact of this `aperture
bias' is discussed in \S\ref{error}.

\subsection{Emission Line Measurement}\label{emlines}

\begin{figure}[h]
\epssz
\plotone{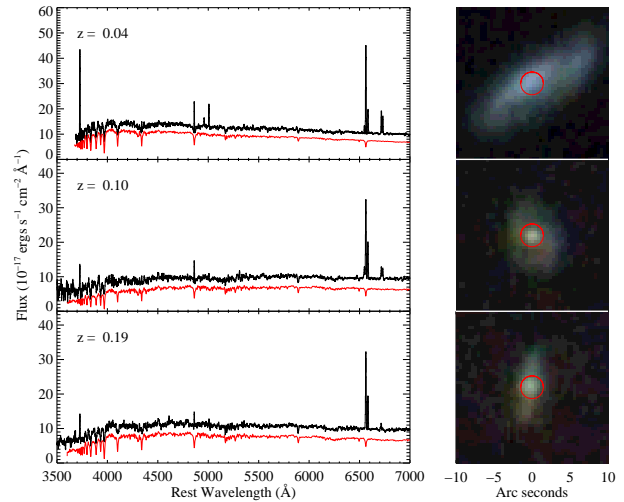}
\caption
{Images and spectra of some typical galaxies in our sample at a range
  of redshifts. The red circle on the images denotes the fiber
  aperture.  The continuum model for each spectrum is shown in red and
  offset slightly for clarity. The continuum model is constructed from
  the \citet{Bruzual_and_Charlot_2003} stellar population synthesis
  models as described in \S\ref{emlines}.  Both observed and model
  spectra have been smoothed by 5 pixels.
\label{ex_gal}}
\end{figure}

The emphasis of this study is on the nebular emission line spectra of
galaxies.  However, we can not disregard the fact that the optical
spectra of galaxies are very rich in stellar \emph{absorption}
features, which can complicate the measurement of nebular emission
lines. In order to maintain speed and flexibility, the SDSS
spectroscopic pipeline performs a very simple estimate of the stellar
continuum using a sliding median.  While this is generally adequate
for strong emission lines, a more sophisticated treatment of the
continuum is required to recover weak features and to properly account
for the stellar Balmer absorption which can reach equivalent widths of
5~\AA\ in some galaxies.  To address this need, we have designed a
special-purpose code optimized for use with SDSS galaxy spectra which
fits a stellar population model to the continuum.  We adopt the basic
assumption that any galaxy star formation history can be approximated
as a sum of discrete bursts.  Our library of template spectra is
composed of single stellar population models generated using the new
population synthesis code of \citet[][hereafter
BC03]{Bruzual_and_Charlot_2003}.  The BC03 models incorporate an
empirical spectral library \citep{Le_Borgne_et_al_2003} with a
wavelength coverage (3200 - 9300 \AA) and spectral resolution
($\sim3$~\AA) which is well matched to that of the SDSS data.  Our
templates include models of ten different ages (0.005, 0.025, 0.1,
0.2, 0.6, 0.9, 1.4, 2.5, 5, 10 Gyr) and three metallicities (1/5
$Z_{\sun}$, $Z_{\sun}$, and 2.5 $Z_{\sun}$).  For each galaxy we
transform the templates to the appropriate redshift and velocity
dispersion and resample them to match the data.  To construct the best
fitting model we perform a non-negative least squares fit with dust
attenuation modeled as an additional free parameter.  In practice, our
ability to simultaneously recover ages and metallicities is strongly
limited by the signal-to-noise of the data.  Hence we model galaxies as
single metallicity populations and select the metallicity which
yields the minimum $\chi^2$.  (While this is not particularly
physical, in practice it is not a bad assumption since the integrated
light of a galaxy tends to be dominated by the light of its most
recent stellar generation.)  The details of the template fitting code
will be presented in Tremonti et al.~ (in prep.).  Figure~\ref{ex_gal}
shows some typical spectra and their continuum models.

After subtracting the best-fitting stellar population model of the
continuum, we remove any remaining residuals (usually of order a few 
percent) with a sliding 200 pixel median, and fit the nebular emission lines.
Since we are interested in recovering very weak nebular features, we adopt a
special strategy: we fit all the emission lines with Gaussians
simultaneously, requiring that all of the Balmer lines
(H$\delta$, H$\gamma$, H$\beta$, and H$\alpha$) have the same line
width and velocity offset, and likewise for the forbidden lines
([\ion{O}{2}]~$\lambda\lambda 3726, 3729$, 
[\ion{O}{3}]~$\lambda\lambda4959, 5007$,
[\ion{N}{2}] $\lambda\lambda6548, 6584$, 
[\ion{S}{2}]~$\lambda\lambda6717, 6731$).  
We are careful to take into account the
wavelength-dependent instrumental resolution of each fiber, which is
measured for each set of observations by the SDSS spectroscopic
pipeline from the arc lamp images.  The virtue of constraining the
line widths and velocity offsets is that it minimizes the number of
free parameters and effectively allows the stronger lines to be used
to help constrain the weaker ones.  Extensive by-eye inspection
suggests that our continuum and line fitting methods work well.

\subsection{The Galaxy Sample}\label{selection}

From our original sample of $\sim211,000$ galaxies, we select a
subsample of star-forming galaxies for nebular analysis.  We start by
imposing a redshift cut of $0.005 < z < 0.25$, and requiring that
$>10$\% of the total galaxy light be observed by the fiber.  Because
stellar mass and metallicity are derived using Bayesian techniques
(see \S\ref{measurements}), it would be possible to select a sample of
objects with accurate masses and metallicities using the likelihood
distributions.  However, for clarity we have opted to define our
sample based on observables (line fluxes, line indices) and later make
some weak cuts based on the derived parameters.  We require galaxies
included in our star-forming sample to have lines of H$\beta$,
H$\alpha$, and [\ion{N}{2}]~$\lambda 6584$ detected at greater than
5$\sigma$.  We do not explicitly constrain the other nebular lines
that we make use of ([\ion{O}{2}]~$\lambda\lambda3726,3729$,
[\ion{O}{3}]~$\lambda5007$, \ion{He}{1}~$\lambda5876$,
[\ion{O}{1}]~$\lambda6300$, [\ion{S}{2}]~$\lambda\lambda6717,6731$)
because our Bayesian analysis takes into account the errors in the
line fluxes.  We note that [\ion{O}{2}] is not measured at $z < 0.03$
due to the blue wavelength cut-off of the spectrographs.  This affects
5\% of our sample but has a negligible impact on our nebular analysis
due to the use of multiple lines (see \S\ref{metallicity}).  We impose
the requirement that the parameters needed for mass determination have
small errors (see \S\ref{mass}).  These criteria are:
$\sigma(\textrm{m}_{z}) < 0.15$ mag, $\sigma(\textrm{H}\delta_{A}) <
2.5$~\AA, and $\sigma[\textrm{D}_n(4000)] < 0.1$.  The combination of
these constraints trims the sample to $\sim82,000$ galaxies, with the
limiting factor typically being the strength of the H$\beta$ line.

We remove galaxies harboring an Active Galactic Nucleus (AGN) from our
sample using the traditional line diagnostic diagram
[\ion{N}{2}]/H$\alpha$ versus [\ion{O}{3}]/H$\beta$
\citep{Baldwin_Phillips_and_Terlevich_1981,
Veilleux_and_Osterbrock_1987}.  The division between star-forming
galaxies and AGN has been calibrated theoretically by
\citet{Kewley_et_al_2001}.  We adopt the slightly modified formula
used by \citet{Kauffmann_et_al_2003c} which provides a more
conservative approach to selecting star-forming galaxies (see their
Fig.~1).  We apply this criterion where we detect [\ion{O}{3}] with
$3\sigma$ sigma significance (and H$\beta$, H$\alpha$, and
[\ion{N}{2}] at 5$\sigma$, as specified earlier.)  To avoid biasing
ourselves against high metallicity galaxies which have intrinsically
weak [\ion{O}{3}], we also include objects with [\ion{O}{3}]~$< 3
\sigma$ and \mbox{$\log$([\ion{N}{2}]/H$\alpha$) $<$ -0.4} in our star
forming sample.  These galaxies comprise $\sim16$\% of our final
sample.  The fraction of emission line galaxies that we reject because
of AGN contamination is $\sim$33\%.

\begin{figure}[h]
\epssz
\plotone{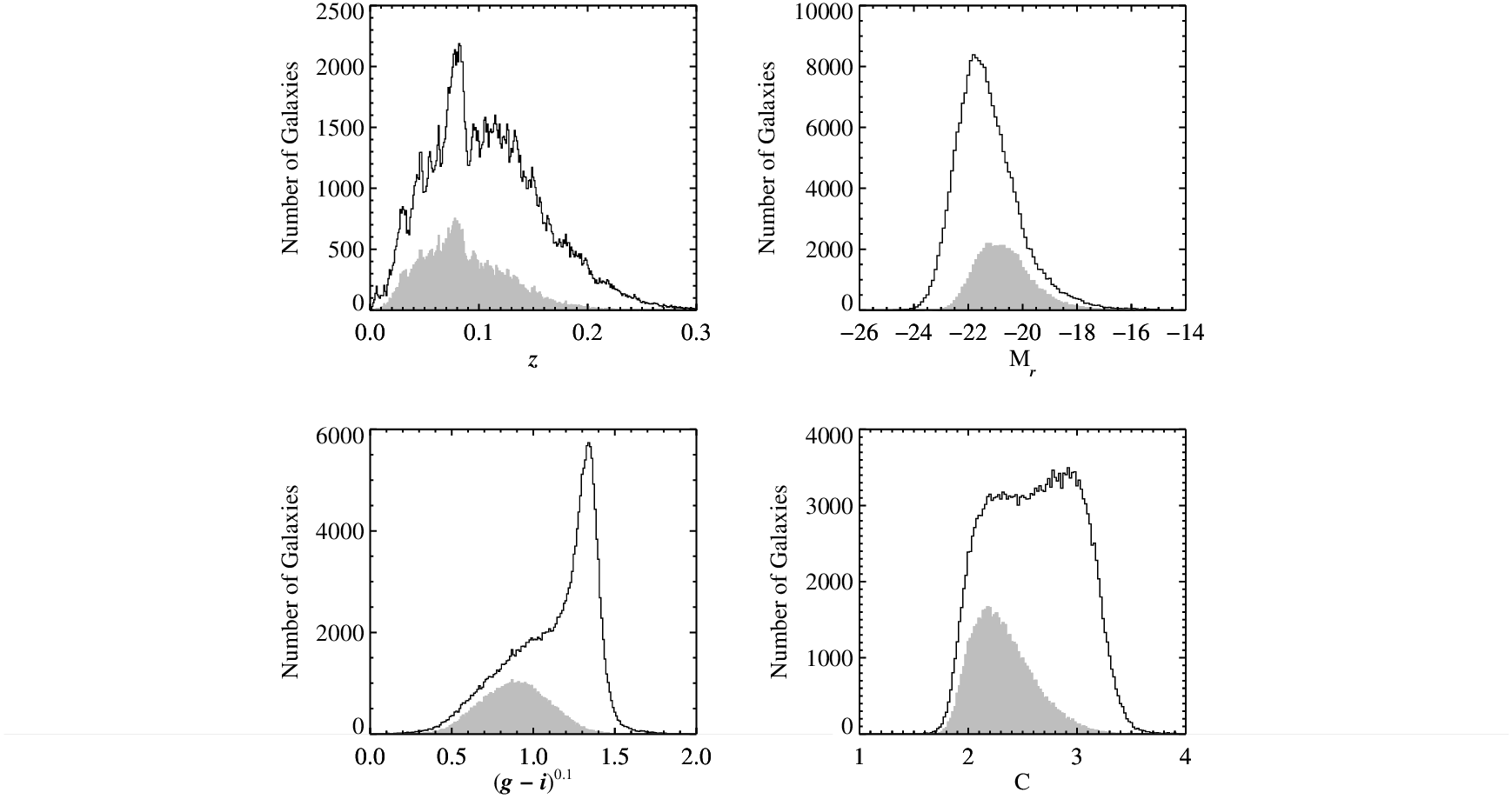}
\caption{Properties of the SDSS main sample and our star-forming
  galaxy sample (gray shaded histograms).  The upper left panel shows
  the redshift distribution.  The upper right panel shows absolute
  $r$-band Petrosian magnitudes.  The lower left panel shows the
  distribution of galaxy ($g-i$) color.  The colors have been
  $k$-corrected to $z=0.1$ following \citet{Blanton_et_al_2003b}.  The
  lower right panel shows the concentration index, $C =
  R_{90}/R_{50}$.  Concentration correlates loosely with Hubble type,
  with $C\sim2.6$ marking the boundary between early- and late-types.
\label{fig_sample}}
\end{figure}

Finally, we require the 1$\sigma$ errors in stellar mass and
metallicity derived from the likelihood distributions to be less than
0.2 dex.  This cut eliminates 3\% of the remaining galaxies, leaving us
with a sample of 53,400 galaxies. Some example galaxies and their
spectra are shown in Figure~\ref{ex_gal}.  The redshift, luminosity,
color, and morphology of our star-forming sample is compared with that
of the DR2 main galaxy sample in Figure~\ref{fig_sample}.  We use the
concentration index, $C$, as a rough proxy for galaxy morphology.  It
is defined as the ratio of the radii enclosing 90\% and 50\% of the
Petrosian $r$-band galaxy light ($C = R_{90}/R_{50}$).
\citet{Shimasaku_et_al_2001} and \citet{Strateva_et_al_2001} have
shown that there is a good correspondence between $C$ and Hubble type,
with $C\sim2.6$ marking the boundary between early- and late-types.
As expected, our emission line sample has bluer colors and more
late-type morphologies than the parent sample.  The redshift
distribution is similar to that of the main survey, and peaks around
$z=0.08$.

\section{The Physical Properties of Galaxies}\label{measurements}

\subsection{Measuring Metallicity}\label{metallicity}

We derive oxygen abundances for our SDSS galaxies from the optical
nebular emission lines.  There are a number of advantages of measuring
metallicity\footnotemark\ from the nebular lines rather than the
stellar absorption features via Lick indices \citep{Worthey_1994}.
First, the S/N in the emission lines can greatly exceed that in the
continuum, with the added advantage that many of the optically
faintest galaxies (dwarfs) exhibit the highest emission line
equivalent widths. Secondly, nebular abundance determination is free
of the uncertainties due to age and $\alpha$-element enhancement which
plague the interpretation of absorption line indices.  Thirdly,
nebular metallicities are easier to interpret in the context of
chemical evolution models because they reflect the present-day metal
abundance rather than the luminosity-weighted average of previous
stellar generations.  The disadvantage is that our analysis is limited
to galaxies with on-going star formation.

\footnotetext{ The word `metallicity' has been used variously in the
literature.  In stellar studies metallicity usually refers to the iron
abundance.  Oxygen has been adopted as the canonical `metal' for ISM
studies because it is the most abundant, it is only weakly depleted on
to dust grains, and it displays strong lines in the optical.
Hereafter we use metallicity to denote the gas-phase oxygen abundance
measured in units of 12 + log(O/H), where the ratio O/H is the
abundance by number of oxygen relative to hydrogen.  Solar metallicity
in these units is 8.69 \citep{Allende_Prieto_et_al_2001} and LMC and
SMC metallicity are 8.4 and 8.0, respectively \citep{Garnett_1999}.}

The strong optical nebular lines of elements other than H and He are
produced by collisionally excited transitions.  The strength of these
lines is the product not only of the heavy element abundance, but the
temperature, density, and ionization state of the nebula, the
attenuation by dust, and the depletion of metals on to dust grains.
In practice, direct determination of element abundances from optical
emission lines tends to be limited by the faintness of the
[\ion{O}{3}]~$\lambda4363$ line which is used to determine the
electron temperature.  When the electron temperature cannot be
measured directly, oxygen abundances can be estimated using
empirically or theoretically calibrated relations between metallicity
and the relative fluxes of various strong optical emission
lines. `Strong-line' abundance calibration was pioneered by
\citet{Alloin_et_al_1979} and \citet{Pagel_et_al_1979}.  The latter
introduced the R$_{23}$ metallicity indicator (R$_{23}$ =
([\ion{O}{2}] $\lambda3727$ + [\ion{O}{3}] $\lambda\lambda4959, 5007$)
/ H$\beta$) which has enjoyed widespread use.  Various other line
ratios have been calibrated as metallicity indicators, including
[\ion{N}{2}]~$\lambda6584$/ H$\alpha$,
[\ion{O}{3}]~$\lambda5007$/[\ion{N}{2}]~$\lambda6584$,
[\ion{N}{2}]~$\lambda6584$/[\ion{O}{2}]~$\lambda3727$, and
([\ion{S}{2}]~$\lambda\lambda6717, 6731$ + [\ion{S}{3}]
$\lambda\lambda90069,9532$)/H$\beta$.  Systematic differences among
the various strong-line calibrations can exceed  0.2 dex. (See
\citealt{Kewley_and_Dopita_2002} for a review).

Here we measure metallicities in a slightly more refined way, using
the approach outlined by \citet{Charlot_et_al_2004}. This consists of
estimating metallicity statistically, based on simultaneous fits of
all the most prominent emission lines ([\ion{O}{2}], H$\beta$,
[\ion{O}{3}], \ion{He}{1}, [\ion{O}{1}], H$\alpha$, [\ion{N}{2}],
[\ion{S}{2}]) with a model designed for the interpretation of
integrated galaxy spectra \citep{Charlot_and_Longhetti_2001}. The
\citet{Charlot_and_Longhetti_2001} model is based on a combination of
the \citet{Bruzual_and_Charlot_1993} and Ferland (1996, version
C90.04) population synthesis and photoionization codes. In this model,
the contributions to the nebular emission by \ion{H}{2} regions and
diffuse ionized gas are combined and described in terms of an
effective (i.e.\ galaxy-averaged) metallicity, ionization parameter,
dust attenuation at 5500~\AA, and dust-to-metal ratio.  The depletion
of heavy elements onto dust grains and the absorption of ionizing
photons by dust are included in a self-consistent way.
\citet{Charlot_and_Longhetti_2001} calibrated the emission-line
properties of their model using the observed [\ion{O}{3}]/H$\beta$,
[\ion{O}{2}]/[\ion{O}{3}], [\ion{S}{2}]/H$\alpha$ and
[\ion{N}{2}]/[\ion{S}{2}] ratios of a representative sample of 92
nearby spiral, irregular, starburst, and \ion{H}{2} galaxies. The
model also accounts well for the properties of these lines in our
sample \citep[][Fig.~5]{Brinchmann_et_al_2004}. We calculate the
likelihood distribution of the metallicity of each galaxy in our
sample, based on comparisons with a large library of models
($\sim2\times10^5$) corresponding to different assumptions about the
effective gas parameters.  We adopt the median of this distribution as
our best estimate of the galaxy metallicity; the width of the
likelihood distribution provides a measure of the error. The median
1$\sigma$ error for our sample of star-forming galaxies is 0.03 dex.
We note that this does not include systematic error.  Because our
metallicities are discreetly sampled, we have added small random
offsets for display purposes. These have a Gaussian distribution with
$\sigma$ = 0.02 dex.

Our models implicitly assume that the nebular gas in a galaxy can be
characterized by a set of galaxy-averaged physical properties.  An
important issue is whether truly representative abundances can be
derived for spatially integrated galaxy spectra in the presence of
radial variations in temperature, ionization, metallicity, and dust
extinction \citep{Vila-Costas_and_Edmunds_1992,
Zaritsky_Kennicutt_and_Huchra_1994, Martin_and_Roy_1994,
Van_Zee_et_al_1998}.  \citet{Kobulnicky_Kennicutt_and_Pizagno_1999}
addressed this issue by comparing \ion{H}{2} region abundances at
particular disk radii with abundances measured from integrated
spectra.  They found excellent agreement ($\pm 0.05$ dex) when
metallicities were based on R$_{23}$.  However, the study is limited
in one respect: they use weighted averages of the \ion{H}{2} region
spectra as proxies for true integrated spectra, thereby ignoring the
contribution of the diffuse ionized gas which can be responsible for
25 - 70\% of the Balmer line emission \citep{Martin_1997,
Zurita_Rozas_and_Beckman_2000, Thilker_et_al_2002}.  Hence further
study is needed to quantify the systematics associated with true
integrated abundances.

\begin{figure}[h]
\epssz
\plotone{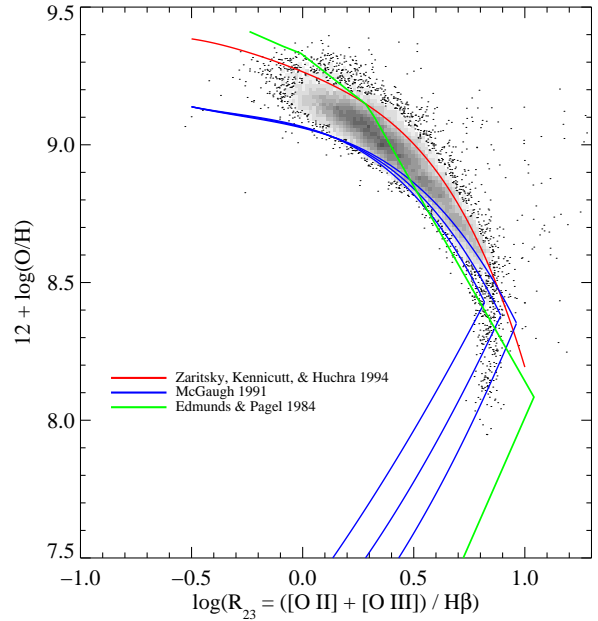}
\caption
{Comparison of the relation between metallicity and the line ratio
 R$_{23}$ = ([\ion{O}{2}] + [\ion{O}{3}] / H$\beta$).  In this and
 subsequent figures we represent the SDSS data using a combination of
 a two dimensional histogram and plotted points.  We histogram the
 data where more than 5 data points fall in an individual pixel, and
 plot it as individual points otherwise. The histograms have been
 square-root scaled for better visibility.  Metallicities for the SDSS
 data have been derived using statistical methods described in the
 text.  The blue line shows the theoretical calibration of
 \citet{McGaugh_1991} (as reported in
 \citet{Kobulnicky_Kennicutt_and_Pizagno_1999}) for three
 representative values of [\ion{O}{3}]/[\ion{O}{2}].  The green line
 shows the empirical calibration of \citet{Edmunds_and_Pagel_1984}, and the
 red line shows the semi-empirical calibration of
 \citet{Zaritsky_Kennicutt_and_Huchra_1994}, itself the average of 3
 previous calibrations.
\label{r23}}
\end{figure}

To place our metallicity measurements in the context of previous work,
we show the relation between our derived metallicities and the line
ratio R$_{23}$ = ([\ion{O}{2}] + \ion{O}{3})/H$\beta$ in
Figure~\ref{r23}.  For comparison we show the theoretical relation
between R$_{23}$ and metallicity derived by
\citet{McGaugh_1991}\footnotemark, the empirical relation defined by
\citet{Edmunds_and_Pagel_1984}, and the semi-empirical relation of
\citet{Zaritsky_Kennicutt_and_Huchra_1994}, itself an average of 3
previous calibrations \citep{Dopita_and_Evans_1986,
Edmunds_and_Pagel_1984, McCall_et_al_1985}.  These analytic
R$_{23}$-metallicity relations roughly bracket the range of
metallicities that we derive, showing that our measurements are in
line with previous strong-line calibrations.  However, recent work
suggests that strong-line methods may overestimate oxygen abundances
\emph{systematically} by as much as a factor of two
\citep{Kennicutt_Bresolin_and_Garnett_2003}.

\footnotetext{We use the analytic formula for the relation between 
R$_{23}$ and metallicity reported in 
\citet{Kobulnicky_Kennicutt_and_Pizagno_1999}.} 

The strength of our Bayesian metallicity estimates is that they make
use of all of the available nebular lines rather than relying on a
small subset of them, as is the case for R$_{23}$.  However, full
spectral modeling is not always possible or practical. Hence we
provide an analytical fit to the R$_{23}$--metallicity relation shown
in Figure~\ref{r23}:
\begin{equation}
12+\log\textrm{(O/H)} = 9.185 - 0.313x - 0.264x^2 - 0.321x^3,
\end{equation}
where $x \equiv \log$R$_{23}$.  This formula is valid for the upper
branch of the double-valued R$_{23}$-abundance relation.  The lower
branch is not very well sampled by our data. We measure metallicities
below 12+log(O/H)=8.5 for 940 galaxies. However less than 1/3 of these
can be plotted in Figure~\ref{r23} because most are at such low
redshift that [\ion{O}{2}]~$\lambda3727$ is not in our spectroscopic
bandpass.  While galaxies lacking [\ion{O}{2}] have larger errors
associated with their derived abundances, they show trends between
abundance and luminosity that are generally consistent with the rest
of the sample.

\subsection{Measuring Stellar Mass}\label{mass}

The stellar mass of a galaxy cannot be inferred directly from its
optical luminosity because the stellar mass-to-light ratio depends
strongly on the galaxy's star formation history and metallicity.
Optical colors have been widely used to estimate M/L ratios
\citep[e.g.][]{Bell_and_de_Jong_2001, Brinchmann_and_Ellis_2000}.
However, errors in the derived M/L are known to result if galaxies
have formed a substantial fraction ($>10\%)$ of their stars in a
recent burst. This caveat is particularly problematic given our sample
of actively star-forming galaxies.  We therefore adopt the method of
\citet{Kauffmann_et_al_2003a} for deriving stellar masses.  This
method relies on spectral indicators of the stellar age and the
fraction of stars formed in recent bursts.  As our present sample is
actually an extension of Kauffmann's, we provide only a cursory
description of our methods here.

We use the $z$-band ($\lambda\sim8900$~\AA) magnitude to characterize
the galaxy luminosity because it is less sensitive than the bluer
bands to extinction and the age of the stellar population.  However,
even in this near-infrared passband the M/L ratio is not a constant,
but can vary by up to a factor of 10 when a full range of star
formation histories is considered \citep[][see
Fig.~4]{Kauffmann_et_al_2003a}.  Constraints on the star formation
history are provided by the spectral indices D$_n$(4000) and
H$\delta_{A}$, which measure the 4000~\AA\ break and the stellar
Balmer absorption.  The location of a galaxy in the D$_n$(4000)--
H$\delta_A$ plane is insensitive to reddening, and only weakly
dependent on metallicity; however, it is a powerful diagnostic of
whether the galaxy has been forming stars continuously or in bursts
over the past 1 - 2 Gyr.  We assign stellar M/L ratios to our galaxies
by using a Bayesian analysis to associate the observed D$_n$(4000) and
H$\delta_A$ values with a model drawn from a large library of Monte
Carlo realizations of galaxies with different star formation histories
and metallicities.  A comparison with broad-band photometry then
yields estimates of the dust attenuation and the stellar mass.  Errors
in our masses are estimated from the width of the likelihood
distribution.  The median 1$\sigma$ error for our sample of
star-forming galaxies is 0.09~dex.  Our masses assume a
\citet{Kroupa_2001} Initial Mass Function (IMF).

\section{The Luminosity--Metallicity Relationship}\label{lummetal}

Because of the relative difficulty of measuring the stellar mass of
galaxies, most previous work has focused on the relationship between
galaxy luminosity and metallicity. Following the pioneering work of
\citet{Lequeux_et_al_1979}, \citet{Skillman_et_al_1989} established a
luminosity--metallicity relation for irregular galaxies which was
confirmed by \citet{Richer_and_McCall_1995} and extended to spiral
galaxies by \citet{Garnett_and_Shields_1987},
\citet{Vila-Costas_and_Edmunds_1992},
\citet{Zaritsky_Kennicutt_and_Huchra_1994}, and
\citet{Kobulnicky_and_Zaritsky_1999}. Recently attempts have been made
to provide better comparison samples for high redshift by studying the
luminosity--metallicity relation in starburst nuclei
\citep{Coziol_et_al_1997}, and UV-selected \citep{Contini_et_al_2002}
and emission line-selected samples \citep{Melbourne_and_Salzer_2002}.

It is instructive to compare the luminosity--metallicity relation of
our sample with previous work, both to validate the consistency of our
measurements and to elucidate the effects of our sample selection.
Although redder bandpasses are less sensitive to the effects of dust
obscuration and recent starbursts, historical precedent mandates that
the luminosity--metallicity relationship be presented in terms of
absolute B magnitude.  Fortunately, the SDSS $g$ band is fairly
similar to the Johnson B band.  We adopt the transformation given in
\citet{Smith_et_al_2002} and $k$-correct the magnitudes to $z=0$ using
the \texttt{kcorrect} code of \citet{Blanton_et_al_2003b}.  We correct
our measured blue luminosities for galactic foreground extinction
\citep{Schlegel_Finkbeiner_and_Davis_1998} and for inclination
dependent intrinsic attenuation \citep{Tully_et_al_1998}. The latter
implies a correction to a face-on orientation, but does \emph{not}
account for the intrinsic attenuation in a face-on system.  The median
correction is $A_B^{inc} = 0.3$~mag.

\begin{figure}[h]
\epssz
\plotone{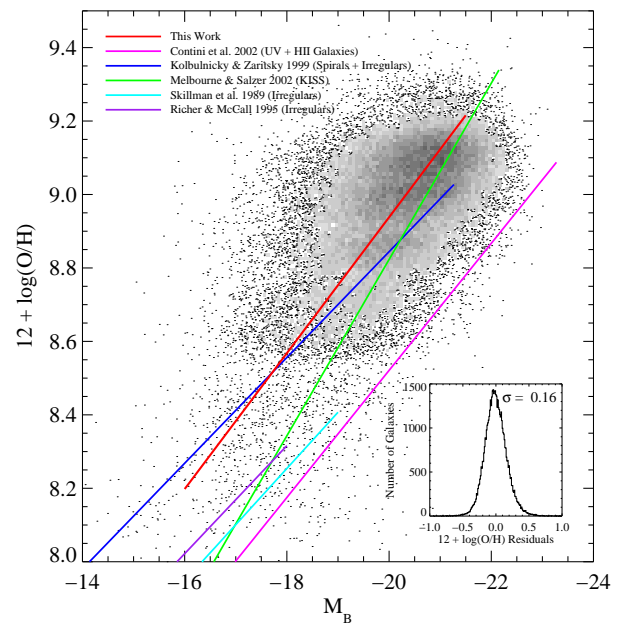}
\caption
{The luminosity--metallicity relation for SDSS galaxies and various
galaxy samples drawn from the literature (see legend).  All of the B
magnitudes have been corrected to $H_0 = 70$~km~s$^{-1}$~Mpc$^{-1}$,
but we have made no attempt to homogenize the metallicity
measurements.  The red line represents the linear least squares
bisector fit to the SDSS data.  The inset plot shows the residuals of
the fit.
\label{lumoh}}
\end{figure}

Figure~\ref{lumoh} shows the luminosity--metallicity relation of the
SDSS galaxies and various samples from the literature.  We have
corrected the published B-band luminosities to our adopted value of
$H_0$ where appropriate, but we have made no attempt to homogenize the
metallicities to account for the different calibrations used.  In
comparing SDSS data with other samples, it is important to consider
the possible aperture bias.  We expect our metallicity measurements to
slightly overestimate the true global abundances, as discussed further
in \S\ref{error}.  However, in view of the different metallicity
calibrations used, the magnitude of this offset is probably not the
dominant systematic effect.  The SDSS data show generally good
agreement with other local samples of galaxies, albeit with a paucity
of dwarfs.
 
In keeping with previous work, we fit the luminosity--metallicity
relation using a linear least squares technique.  The measurement
errors in luminosity ($\pm 0.01$~dex) and metallicity ($\pm 0.1$~dex)
are small compared to the observed scatter.  The traditional method of
estimating the underlying functional relation in this case is to use
the linear bisector --- the line which bisects the ordinary
least-squares regression of $X vs.\ Y$ and $Y vs.\ X$
\citep{Isobe_et_al_1990}.  The luminosity--metallicity relation for
our galaxies fitted in this manner is
\begin{equation}
12+\log(\textrm{O/H}) = -0.185 (\pm 0.001) \textrm{M}_{B} + 5.238 (\pm 0.018). 
\end{equation}
Our measured slope is intermediate between that found by
\citet{Kobulnicky_and_Zaritsky_1999} for local irregulars and spirals
and that measured by \citet{Melbourne_and_Salzer_2002} for a subset of
galaxies from the KPNO International Spectroscopic Survey (KISS).  In
inter-comparing data sets it is important to note that different
authors (including the two mentioned above) have adopted different
regression methods. Different regression methods calculate
intrinsically different properties of the data and \emph{are not
directly comparable} (see \citet{Isobe_et_al_1990} for a full
discussion).  Figure~\ref{lumoh} therefore elucidates the differences
in the measured luminosity--metallicity relation which are the result
of sample selection, metallicity calibration, and data
characterization (fitting).  All of these factors need to be carefully
considered when looking for evidence of evolution in the
luminosity--metallicity relation.

\begin{figure}[h]
\epssz
\plotone{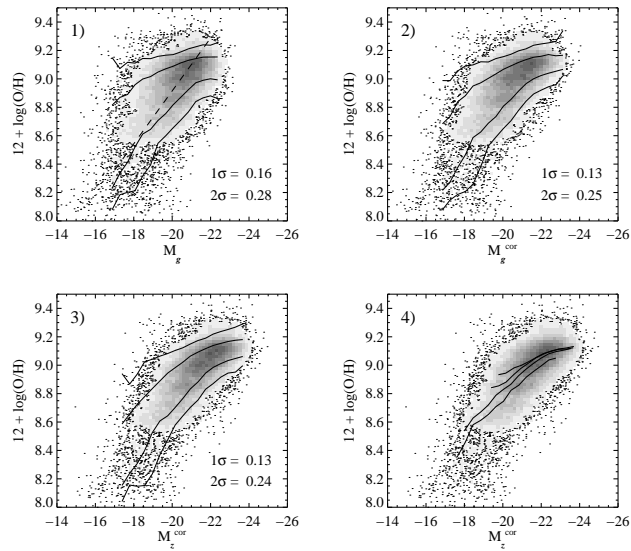}
\caption {The luminosity--metallicity relation of SDSS galaxies in the
  $g$ and $z$-bands. In the first panel we have corrected M$_{g}$ to
  face-on orientation, but we have not corrected for internal
  attenuation.  In panels 2-4 we correct M$_{g}$ and M$_{z}$ for
  internal attenuation, assuming that the stars experience 1/3 of the
  reddening measured in the gas.  The solid black contours in panels
  1-3 enclose 68 and 95\% of data with statistics computed in bins of
  0.4 mag in luminosity.  The median half-width of the distribution is
  listed in the lower right corner.  For comparison, the dashed line
  in the first panel shows the least-squares linear bisector fit to
  the data.  The fourth panel shows the median $z$-band
  luminosity--metallicity relation for galaxies in four bins of
  D$_n$(4000): from bottom to top, 1.0 - 1.2, 1.2 - 1.3, 1.3 - 1.4,
  1.4 - 1.8.  Data for the contours in panels 1 and 3 are given in
  Tables~1 and~2.
\label{lumz_gz}}
\end{figure}

At fixed metallicity, the SDSS galaxies span more than 4 magnitudes in
blue luminosity.  The brightest galaxies are consistent with the
luminosity--metallicity relation found by \citet{Contini_et_al_2002}
for a sample of UV-selected galaxies (pink line in Fig.~\ref{lumoh}).
Contini et al.\ postulate that these galaxies are offset from the
luminosity--metallicity relation of normal spirals due to the low M/L
ratio of their newly formed stellar populations.  This suggests a
physical origin for the scatter, an idea that we explore further in
Figure \ref{lumz_gz}.  To reduce sources of measurement error, we
switch to using the luminosity in the native SDSS passbands,
$k$-corrected to the median redshift, $z\sim0.1$.  To quantify the
influence of dust, we correct the galaxy luminosities for intrinsic
attenuation using the attenuation curve of
\citet{Charlot_and_Fall_2000} and assuming that the stars experience
1/3 of the attenuation measured in the nebular gas.  The nebular
attenuation is determined simultaneously with the metallicity assuming
a metallicity-dependent Case~B H$\alpha$/H$\beta$ ratio
\citep{Brinchmann_et_al_2004, Charlot_et_al_2004}.  We assume that the
correction for intrinsic attenuation automatically accounts for any
inclination-dependent effects.

In Figure~\ref{lumz_gz} we systematically examine the impact of dust
and M/L variations on the luminosity--metallicity relation by adopting
different measures of galaxy luminosity: 1) the absolute $g$-band
magnitude corrected for inclination-dependent attenuation following
\citet{Tully_et_al_1998}; 2) the absolute $g$-band magnitude corrected
for intrinsic attenuation, as described above; 3) and the absolute
$z$-band magnitude corrected for intrinsic attenuation.  In panels 1-3
of Figure~\ref{lumz_gz} we indicate the \emph{distribution} of
metallicity at a given luminosity by displaying the contours which
enclose 68 and 95\% of the data in bins of 0.4 mag.  The contours
provide a non-parametric description of the distribution which is
unbiased as long as the errors in luminosity are small relative to our
adopted bin-size.  For comparison, we also show the traditional
least-squares linear bisector fit to the data in panel~1 (12+log(O/H)
= $-0.186(\pm0.001)M_g + 5.195(\pm0.018)$).  Because we do not know a
priori the true functional form of the luminosity--metallicity
relation, we focus on the contours.  Comparison of the first two
panels of Figure~\ref{lumz_gz} shows that correcting the luminosity
for attenuation reduces the scatter and flattens the
luminosity--metallicity relation at high mass.  This trend is even
more pronounced when the extinction corrected $z$-band magnitude is
used (panel~3).  Because the $z$-band is less sensitive to dust and
recent starbursts, the range of M/L ratios is smaller, and the scatter
is reduced by $\sim20$\% compared to the uncorrected $g$-band.
However, even in the $z$-band, M/L ratios can vary by factors of a
few.  This effect is illustrated in panel~4 where we plot the median
$z$-band luminosity--metallicity relation for galaxies in four bins of
D$_n$(4000).  As discussed in \citet{Kauffmann_et_al_2003a},
D$_n$(4000) is a good measure of the mean stellar age of the
population.  Our interpretation of panel~4 is that at fixed
metallicity, galaxies with lower D$_n$(4000) are shifted to brighter
magnitudes because of the lower M/L ratios of their young stellar
populations.  This confirms our intuition that the underlying physical
correlation is between stellar mass and metallicity.

\section{The Mass--Metallicity Relationship}\label{massmetal}

\begin{figure}[h]
\epssz
\plotone{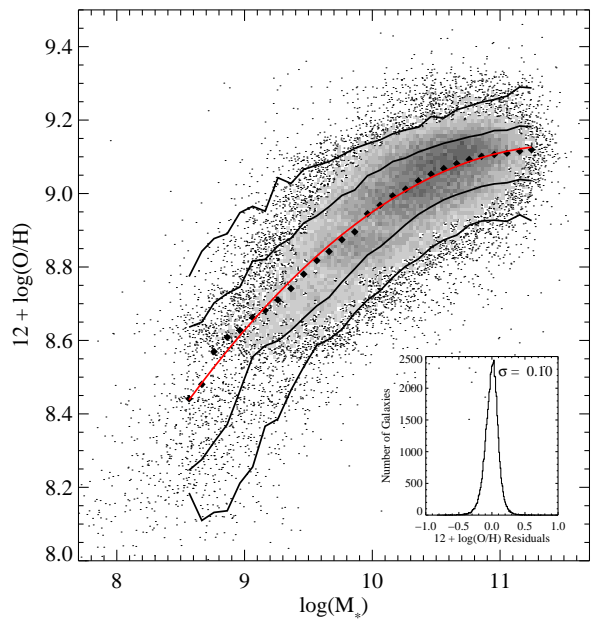}
\caption
{The relation between stellar mass, in units of solar masses, and
gas-phase oxygen abundance for $\sim$53,400 star-forming galaxies in
the SDSS.  The large black points represent the median in bins of 0.1
dex in mass which include at least 100 data points.  The solid lines
are the contours which enclose 68\% and 95\% of the data.  The red
line shows a polynomial fit to the data.  The inset plot shows the
residuals of the fit.  Data for the contours are given in Table~3.
\label{massoh}}
\end{figure}

With our new prescriptions for measuring stellar mass and gas-phase
metallicity it is now possible to examine the mass--metallicity
relationship of our sample of SDSS star-forming galaxies.  Figure
\ref{massoh} shows that a striking correlation is observed, extending
over 3 decades in stellar mass and a factor of 10 in metallicity.  The
correlation is roughly linear from $10^{8.5}$~M$_{\sun}$ to
$10^{10.5}$~M$_{\sun}$ after which a gradual flattening occurs.  Most
remarkable of all is the tightness of the correlation: the 1$\sigma$
spread of the data about the median is $\pm0.10$~dex, with only a
handful of extreme outliers present.  The relationship is well fitted
by a polynomial of the form:
\begin{equation}
\textrm{12+log(O/H)} = -1.492 + 1.847 (\log \textrm{M}_{*}) 
 - 0.08026 (\log \textrm{M}_{*})^2
\end{equation}              
where M$_{*}$ represents the stellar mass in units of solar masses.  
This equation is valid
over the range $8.5 < \log\textrm{M}_{*} < 11.5$.

The principal difference between the mass--metallicity and
luminosity--metallicity relations is the more pronounced turnover seen
at high metallicity when mass is used as the independent variable. In
Figure~\ref{lumz_gz} we demonstrated that both dust and M/L variations
act to smear out the turnover when luminosity is used in place of
mass.  Accordingly, the scatter in the uncorrected $g$-band
luminosity--metallicity relation is $\sim$50\% higher than the scatter
in the mass--metallicity relation.

We emphasize that the turnover in the mass--metallicity relation is
not an artifact of our metallicity calibration. As shown in
Figure~\ref{r23} our metallicity estimates are consistent with other
strong-line calibrations.  If we adopt the R$_{23}$ metallicity
calibration of \citet{Zaritsky_Kennicutt_and_Huchra_1994}, the shape
of the mass--metallicity relation is well reproduced, but the scatter
increases by $\sim15$\%.  Use of the \citet{McGaugh_1991} R$_{23}$
metallicity calibration results in a \emph{more} pronounced turnover,
due to the flatter relation between R$_{23}$ and log(O/H).

The shape of the mass--metallicity relation is also robust with
respect to our choice of S/N cuts.  The S/N in the H$\beta$ line is
generally the parameter that drives our galaxy selection.  The main
concern is that our requirement of (S/N)$_{H\beta} > 5$ may bias us
toward massive galaxies with abnormally high gas fractions.  However,
at fixed equivalent width (EW) the distribution of S/N values is
broad.  Galaxies with H$\beta$ EW $< 5$~\AA\ have a median
(S/N)$_{H\beta}$ of 11, and more than 10\% of them have
(S/N)$_{H\beta} > 20$.  Thus our cuts on S/N are not imposing
stringent cuts on EW or gas fraction.  We have also done the
straightforward experiment of altering the S/N cuts on all our lines
and re-fitting the mass--metallicity relation: its detailed shape is
insensitive to factor of 2 changes in our S/N cuts.

\begin{figure}[h]
\epssz
\plotone{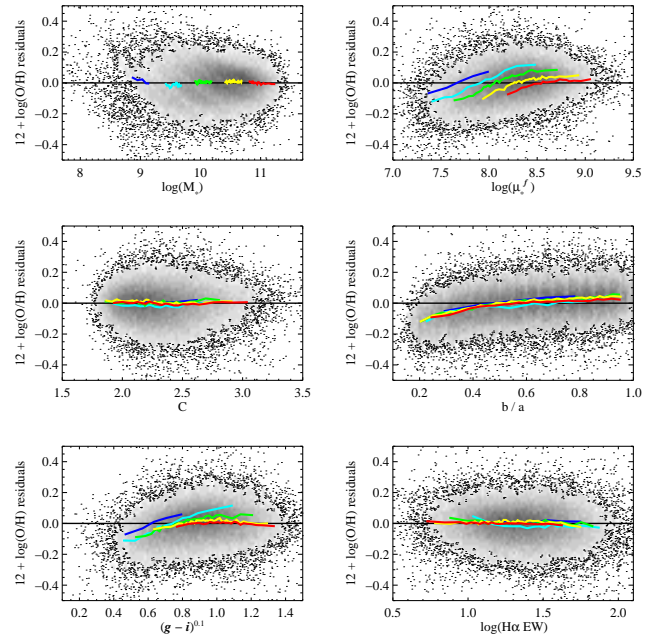}
\caption
{Correlation of the residuals of the mass--metallicity relation with
 various galaxy physical parameters: $M_{*}$, the stellar mass in
 units of M$_{\sun}$; $\mu_{*}^{f}$, the local surface mass density
 measured within the fiber aperture in units of M$_{\sun}$~kpc$^{-2}$;
 $C$, the concentration parameter, a proxy for galaxy morphology;
 $b/a$, the ratio of the galaxy minor and major axes, a proxy for
 inclination; ($g-i$) color $k$-corrected to $z$=0.1; and H$\alpha$
 equivalent width in \AA.  The colored lines show the median relation
 in 5 different mass bins.
\label{resid}}
\end{figure}

The most remarkable feature of the stellar mass--metallicity relation
is the tightness of the correlation -- the 1$\sigma$ spread of the
data about the median ranges from 0.20~dex at low mass to 0.07 dex at
high mass. For comparison, the median errors in mass and metallicity
are 0.09~dex and 0.03~dex respectively.  Adding these appropriately,
we find that roughly half of the spread in the mass--metallicity
relation can be attributed to observational error.  With our large
statistical sample we can test whether there is a physical origin for
the scatter, even if we have underestimated our measurement errors.
In Figure~\ref{resid} we show the residuals from our fit to the
mass--metallicity relation as a function of various galaxy physical
properties.  

There is a clear tendency for the metallicity residuals to correlate
with local surface mass density measured within the fiber aperture.
Analogous trends were identified by \citet[]{Bell_and_de_Jong_2000}
using resolved photometry of a sample of nearby disk galaxies. They
demonstrated that mean stellar metallicity and stellar age correlate
with local K-band surface density
and concluded that surface density plays a strong role in shaping both
the local and global star formation history of a galaxy.
\citet{Kauffmann_et_al_2003b} showed that the star formation histories
of low mass galaxies are more fundamentally related to surface mass
density than to stellar mass.  These findings suggest that at fixed
mass, galaxies with higher surface densities have transformed more of
the available gas into stars, thereby raising the gas-phase
metallicity.  This interpretation is borne out by the fact that the
residuals also correlate somewhat with galaxy color. But curiously,
the same trend is not seen with H$\alpha$ EW or galaxy morphology as
measured by the concentration index.  This leaves us with a somewhat
contradictory picture since these quantities are expected to relate to
a galaxy's gas content and general evolutionary state.

The other pronounced correlation seen in Figure~\ref{resid} is between
metallicity and inclination, as measured by the ratio of the galaxy
minor and major axes.  At fixed mass, fully edge-on galaxies
(b/a$\sim0.2$) are more metal poor than face-on galaxies by 0.19 dex.
This trend is easily explained because more of the disk is seen in
projection in our fiber aperture at higher inclinations.  The
integrated metallicity decreases as larger galaxy radii are probed
because the outer regions of disks tend to be more metal-poor, as well
as more gas rich and less extincted.

\section{The Origin of the Mass--Metallicity Relation}\label{mzorigin}

The existence of a tight ($\pm0.1$~dex) correlation between stellar
mass and metallicity reflects the fundamental role that galaxy mass
plays in galactic chemical evolution.  However, it is not a priori
clear whether this sequence is one of enrichment or of depletion.
Simply put, if more massive galaxies form fractionally more stars in a
Hubble time than their low-mass counterparts, then the observed
mass--metallicity relation represents a sequence in astration.
However, if galaxies form similar fractions of stars, then the
relation could imply that metals are selectively lost from galaxies
with small potential wells via galactic winds.

Both of these ideas have a long history in the astronomical
literature.  It is well known that there are systematic trends in the
star formation history of galaxies along the Hubble sequence
\citep[e.g.][]{Roberts_and_Haynes_1994}. Recent observations have
confirmed that gas mass fractions decrease with increasing stellar
mass \citep{McGaugh_and_de_Block_1997, Bell_and_de_Jong_2000,
Boselli_et_al_2001}, a trend that seems to suggest that low mass
galaxies are un-enriched rather than depleted.  However, observations
of starbursts have revealed the nearly ubiquitous presence of galactic
winds \citep{Heckman_2002}, while studies of X-ray bright clusters
have demonstrated the presence of copious metals in the intracluster
medium \citep{Gibson_et_al_1997}, and absorption line studies have
revealed metals in the intergalactic medium
\citep{Ellison_et_al_2000}.  Given the existence of these
contradictory pieces of information, anecdotal arguments are of little
use: a direct test of the origin of the mass--metallicity relation is
required.

The effects of astration and mass loss on the relationship between
mass and metallicity can be disentangled in principle if information
on the relative mass of the gas and stars is available.  Simple closed
box chemical evolution models predict that metallicity ($Z$) is a
simple function of the stellar yield, $y$, and of the gas mass
fraction, $\mu_{gas}$:

\begin{equation}
Z = y \ln(\mu_{gas}^{-1}) 
\end{equation}

Following \citet{Garnett_2002}, we assume that the stellar yield is
constant. Metallicity is then straightforwardly related to the gas
mass fraction if the tenets of the simple model apply -- namely that
there are no gas inflows or outflows.  We can invert Equation~1 to
define the `effective yield', which can be computed from the observed
metallicity and gas mass fraction.  When the simple model applies, the
effective yield will equal the true yield, independent of galaxy mass.
This straightforward observational test was performed by
\citet{Garnett_2002} using a sample of 44 nearby spiral and irregular
galaxies.  Garnett found the effective yield to be constant for
galaxies with rotational velocities in excess of $\sim150$~km~s$^{-1}$
and to decline by a factor of $\sim15$ below this threshold.
 
While our SDSS dataset is much larger than the nearby sample of
Garnett, we do not have direct information about the \ion{H}{1} and
H$_2$ gas content of our galaxies.  However, we do have indirect
information, by virtue of the fact that all of our sample galaxies are
actively forming stars.  To estimate the gas mass, we invoke another
well known empirical correlation -- the Schmidt star formation law
\citep{Schmidt_1959,Kennicutt_1998} which relates the star formation
surface density to the gas surface density.

For each of our galaxies we calculate the star formation rate (SFR) in
the fiber aperture from the attenuation-corrected H$\alpha$ luminosity
following \citet{Brinchmann_et_al_2004}.  
We multiply our SFRs by a factor of 1.5 to convert from a
\citet{Kroupa_2001} IMF to the Salpeter IMF used by
\citet{Kennicutt_1998}.  Our SDSS galaxies have star formation surface
densities which are within a factor of 10 of $\Sigma_{SFR} =
0.3$~M$_{\sun}$~yr$^{-1}$~kpc$^{-2}$, exactly the range found by
\citet{Kennicutt_1998} for the central regions of normal disk
galaxies.  We convert star formation surface density to surface gas
mass density, $\Sigma_{gas}$, by inverting the composite Schmidt law
of \citet{Kennicutt_1998},
\begin{equation}
\Sigma_{SFR} = 1.6 \times 10^{-4}  \left(\frac{\Sigma_{gas}}
                 {1 \textrm{M}_{\sun} \textrm{~pc}^{-2}}\right)^{1.4}
                 \textrm{M}_{\sun}~\textrm{yr}^{-1} \textrm{kpc}^{-2}. 
\end{equation}
(Note that the numerical coefficient has been adjusted to include
helium in $\Sigma_{gas}$.)  Combining our spectroscopically derived
M/L ratio with a measurement of the $z$-band surface brightness in the
fiber aperture, we compute $\Sigma_{star}$, the stellar surface mass
density.  The gas mass fraction is then $\mu_{gas} = \Sigma_{gas} /
(\Sigma_{gas} + \Sigma_{star})$.

\begin{figure}[ht]
\epssz
\plotone{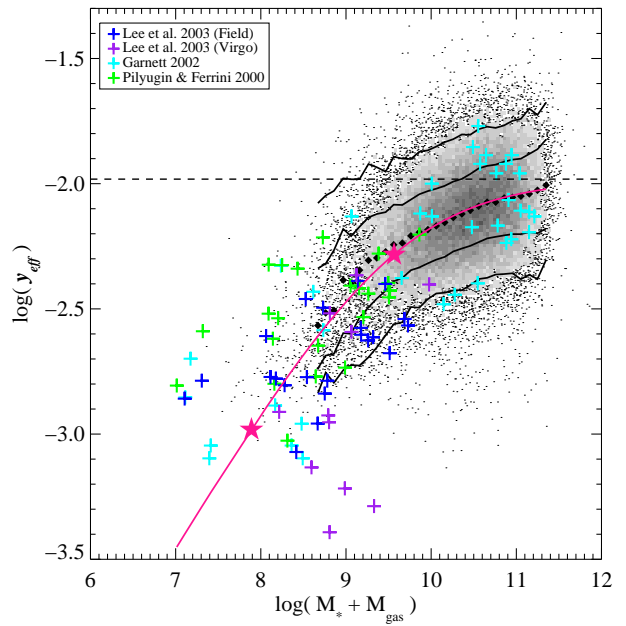}
\caption{Effective yield as a function of total baryonic mass (stellar
+ gas mass) for 53,400 star-forming galaxies in the SDSS.  The large
black points represent the median of the SDSS data in bins of 0.1 dex
in mass which include at least 100 data points.  The solid lines are
the contours which enclose 68\% and 95\% of the data.  The colored
crosses are data from \citealt{Lee_McCall_and_Richer_2003},
\citealt{Garnett_2002}, and \citealt{Pilyugin_and_Ferrini_2000}.  Both
the metallicities and the gas masses used to derive the effective
yield have been computed differently in the SDSS data and the samples
from the literature.  The agreement nevertheless appears quite good.
The pink line is the best fit to the combined dataset assuming the
intrinsic functional from given by Equation~6.  The dashed line
indicates $y_0$, the true yield if no metals are lost, derived from
the fit to the data. The pink stars denote galaxies which have lost
50\% and 90\% of their metals.  Data for the contours are given in
Table~4.
\label{gasmass}}
\end{figure}

In Figure~\ref{gasmass} we plot the effective yield of our SDSS star
forming galaxies as a function of total baryonic (stellar + gas) mass.
Baryonic mass is believed to correlate with dark mass, as evidenced by
the existence of a baryonic `Tully-Fisher' relation
\citep{McGaugh_et_al_2000, Bell_and_de_Jong_2001}.  We are interested
in the dark mass because departures from the `closed box' model might
be expected to correlate with the depth of the galaxy potential well.
Because very few of our SDSS galaxies have masses below
10$^{8.5}$~M$_{\sun}$, we augment our data set with measurements from
\citet{Lee_McCall_and_Richer_2003}, \citet{Garnett_2002}, and
\citet{Pilyugin_and_Ferrini_2000}, \emph{all of which use direct gas
mass measurements}.  We note that the correspondence between the SDSS
data and the samples from the literature is very good in spite of the
fact that the gas masses have been determined in very different ways.

With the addition of the data from the literature it is clear that
galaxies do not evolve as `closed boxes'.  Figure~\ref{gasmass} shows
that baryonic mass and effective yield are highly correlated, with the
effective yield decreasing by a factor of $\sim10$ from the most
massive galaxies to dwarfs.  The correlation is steep at low masses
but begins to flatten around $10^{9.5}$~M$_{\sun}$.  In
\S\ref{discussion} we argue that the relationship between effective
yield and baryonic mass is the consequence of metal loss via galactic
winds.  We fit the combined dataset with a simple function predicated
on this basis.  We assume that the fraction of newly synthesized
metals retained by a galaxy is proportional to the depth of the
galaxy's potential well for low mass galaxies, but asymptotes to unity
for the most massive systems.  The depth of the galaxy potential well
scales approximately as $V_c^2$, where $V_c$ is the galaxy circular
velocity.  We model the retained fraction as $f_{ret} = V_c^2 / (V_c^2
+ V_0^2)$ where $V_0$ is a constant.  ($f_{ret}(V_c < V_0) < 0.5$,
$f_{ret}(V_c \gg V_0)\sim1$.)  We adopt $M_{baryon} \propto V_c^{3.5}$
from the baryonic Tully-Fisher relation of
\citet{Bell_and_de_Jong_2001}.  The effective yield and the baryonic
mass may then be related follows:
\begin{equation}
   y_{eff} = \frac{y_0}{(1 + (M_{0}/M_{baryon})^{0.57})} 
\end{equation}
The free parameters of the fit are $y_0$, the true yield (if no metals
are lost), and $M_0$, the mass at which a galaxy loses 1/2 of its
metals.  We perform the fit to a combination of the SDSS data (the
median value of $y_{eff}$ in bins of 0.1 in $M_{baryon}$) and the data
drawn from the literature, applying weights such that both datasets
contribute equally to the fit.  We find $y_0 = 0.0104$ and $M_{0} =
3.67\times 10^9$~M$_{\sun}$.  This characteristic mass corresponds to a
rotation speed of $V_{c}=85$~km~s$^{-1}$ and, following
\citet{Heckman_et_al_2000}, an escape velocity of $V_{esc} \sim 3 V_c
= 260$~km~s$^{-1}$.  Thus, the escape velocity of a galaxy that has
lost half of its metals is just below the observed terminal velocity
of starburst winds, which are 300 - 900 km~s$^{-1}$
\citep{Heckman_et_al_2000}.

The scatter in the correlation between effective yield and baryonic
mass is $\pm0.15$~dex, 50\% larger than that measured for the
mass--metallicity relation. However, much of the scatter in the SDSS
data can probably be ascribed to our relatively uncertain measurement
of the gas mass. We note that this does not effect the total mass
greatly, as the median gas mass fraction of our sample is $\sim$20\%.
It is interesting to note that galaxies with direct gas mass
measurements show comparable scatter, however the stellar mass
determinations for these galaxies are correspondingly more uncertain.
It is therefore premature to consider a physical origin for the
scatter, although it is quite likely that there may be one, for
instance differences in galaxy environment \citep{Skillman_et_al_1996,
Lee_McCall_and_Richer_2003}.

\section{Sources of Systematic Error}\label{error}

\begin{figure}[ht]
\epssz
\plotone{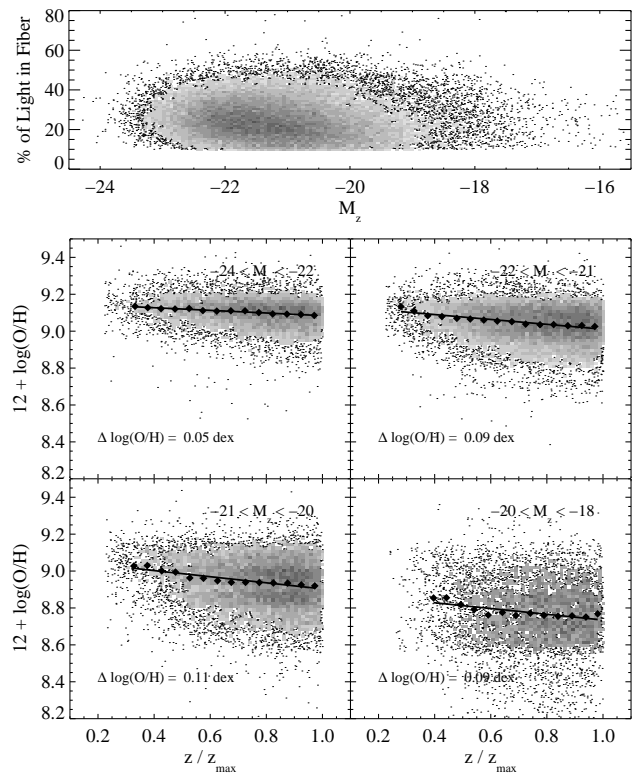}
\caption{A test for aperture bias.  In the top panel the fraction of
$r$-band galaxy light observed by the fiber is plotted as a function
of absolute magnitude. In the lower panel, metallicity is plotted in 4
broad bins of absolute magnitude as a function of relative redshift,
$z/z_{max}$, where $z_{max}$ is the redshift at which a given galaxy
reaches the survey limits.  The large black points represent the
median; the solid line is a linear fit to the points. The maximum
change in metallicity with $z/z_{max}$ (based on the fitted line) is
given in the lower left of each panel.
\label{apbias}}
\end{figure}

Galaxies are known to have radial gradients in their physical
properties \citep[e.g.][]{Bell_and_de_Jong_2000}.  Aperture effects
are therefore a potentially serious concern because strong selection
effects are inherent in magnitude-limited surveys such as the SDSS.
For our sample, the median projected fiber size is $\sim$4.6~kpc in
diameter, with the range extending from 1 to 12~kpc.  Thus while we
are not measuring `nuclear' spectra, we are clearly not measuring
integrated spectra either. The median fraction of galaxy light in the
fiber aperture is 24\%.  Fortunately, because larger galaxies are
brighter, they get selected out to larger distances and observed with
larger projected apertures. The net effect is that the fraction of
galaxy light seen by the fibers is not a strong function of absolute
magnitude, as shown in the top panel of Figure~\ref{apbias}.  Hence,
while aperture effects are important in the absolute sense, they are
unlikely to have a significant effect on trends with galaxy mass.

Because of the magnitude-limited nature of the SDSS, it is not
possible to examine the impact of aperture size simply by comparing
galaxies in narrow bins of absolute magnitude at different redshifts.
Instead we divide our sample into four broad bins of absolute
magnitude and examine trends in galaxy properties as a function of
relative redshift, $z/z_{max}$, where $z_{max}$ is the redshift at
which a given galaxy reaches the spectroscopic survey limits.  The
effect of aperture bias on galaxy M/L ratios was examined as a
function of $z/z_{max}$ by \citet[][see Fig.~18]
{Kauffmann_et_al_2003a} and found to be $\sim0.1$ dex.  The strongest
trends were observed for L$^{*}$ galaxies which presumably have both a
well developed bulge and a disk.  The effect of aperture bias on our
measured metallicity is shown in Figure~\ref{apbias}.  We find a
change of at most -0.11 dex in metallicity as galaxies move from one
edge of the survey to the other, indicating that our metallicity
determinations are only moderately affected by changes in the
projected aperture size. This gradient is in good accord with our
expectations based on the known radial metallicity gradients in spiral
galaxies \citep{Vila-Costas_and_Edmunds_1992,
Zaritsky_Kennicutt_and_Huchra_1994, Martin_and_Roy_1994,
Van_Zee_et_al_1998}.

While it is desirable to correct our metallicity measurements to
reflect true global abundances, it is not possible to derive this
correction directly from the SDSS data because even at the limits of
the survey the fiber aperture only contains about 50\% of a galaxy's
light. Work is under way to obtain the observations necessary for such
an analysis.  In the meantime, we simply note that because the fibers
generally cover the inner few kpc of galaxies, the abundances we
derive are likely to be higher than true integrated abundances.
Recent evidence also suggests that oxygen abundances derived from
`strong lines' methods like ours may overestimate the true abundance
by as much as factor of two
\citep{Kennicutt_Bresolin_and_Garnett_2003}.

Another potential source of error is our indirect method of deriving
the gas mass from the H$\alpha$ luminosity via the global Schmidt law
of \citet{Kennicutt_1998}.  While the impressive correlation between
the surface mass density of gas and the surface density of star
formation holds over 5 orders of magnitude, there is a fair amount of
scatter evident in the narrow regime applicable to the present study
\citep[][see Fig.~6]{Kennicutt_1998}.  Systematic errors are a bigger
concern.  \citet{Wong_and_Blitz_2002} found that application of a
radially-dependent attenuation correction to H$\alpha$ produced a
steeper Schmidt law, with a power law index of 1.7 (as compared to
Kennicutt's 1.4).  However, adopting the steeper Schmidt law in our
analysis produces a relatively small change in the correlation between
baryonic mass and effective yield: log$y_0$ decreases by 0.04 dex, and
log$M_0$ decreases by 0.15 dex.

Beyond the issue of the accuracy of our $\mu_{gas}$ measurement, there
is the larger question of whether the central gas mass fraction is the
\emph{relevant} gas mass fraction.  If $\Sigma_{gas}/\Sigma_{star}$
varies strongly with radius, than our measurements are unlikely to be
representative of the galaxy as a whole.  These concerns are allayed
to some degree by the generally strong correspondence seen between
H$\alpha$ and broad-band optical disk scale lengths
\citep{Hodge_and_Kennicutt_1983, Kennicutt_1989}.  A more fundamental
question, however, is how the products of star formation are
distributed in galaxies -- in other words, how well mixed is the
interstellar medium?  The mere existence of radial metallicity
gradients in spirals suggests that mixing timescales are generally
long.  In this case, local abundances reflect local astration levels,
and hence the relevant gas mass fraction is the local one.  A final
issue is the existence of cold gas external to the stellar disk in
some galaxies.  It is unclear whether this gas is relevant to our
calculation, since it does not strongly participate in the star
formation.  However, if the outlying gas viscously mixes with gas in
the inner disk it would affect the observed metallicity, and should be
included in the gas budget.  In any case, since we have no information
on the extended gas in our galaxies, we do not consider it at present,
except to note that its inclusion would raise our measured value of
the effective yield.  For galaxies with baryonic masses of
$10^7~$M$_{\sun}$ the gas mass would need to increase by a factor of
$\sim11$ to erase the signatures of metal loss.

While the sources of systematic error cannot be ignored, we believe
the magnitude of these errors to be small, on the order of a few
tenths of a dex. In view of the moderately large dynamic range of our
results -- a factor of $\sim10$ decrease in both metallicity and
effective yield -- our general result, the preferential loss of metals
from galaxies with small potential wells, appears very robust.  Future
observations which spatially resolve the distribution of gas and
metals in galaxies will prove key to understanding many of the
relevant systematic effects.

\section{Summary \& Discussion}\label{discussion}

We have coupled advanced techniques for deriving stellar masses and
gas-phase metallicities from optical spectroscopy and photometry with
the statistical power provided by the SDSS.  For a sample of 53,400
star-forming galaxies at $z\sim0.1$, we find a tight correlation
($\pm0.1$~dex) between stellar mass and the gas-phase oxygen
abundance, which extends over 3 orders of magnitude in stellar mass
and a factor of 10 in oxygen abundance.  We use indirect estimates of
the gas mass fraction based on our measured H$\alpha$ luminosity to
estimate the effective yield.  Simple closed box chemical evolution
models predict that the effective yield is constant. Combining our
SDSS data with data from the literature, we find evidence that the
effective yield decreases by a factor of $\sim10$ from the most
massive galaxies to dwarfs.

The most straightforward interpretation of the correlation between
baryonic mass and effective yield is the selective loss of metals from
galaxies with shallow potential wells via galactic winds, an idea
first introduced by \citet{Larson_1974}.  Evidence for galactic
outflows has steadily accumulated from observations at optical
\citep{Lehnert_and_Heckman_1996, Heckman_et_al_2000}, X-ray,
\citep{Dahlem_et_al_1998, Martin_et_al_2002}, and ultraviolet
wavelengths \citep{Heckman_and_Leitherer_1997, Pettini_et_al_2000}.
However, comparatively little is known about the eventual \emph{fate}
of the wind material --- whether it escapes from the galaxy potential,
or cools and eventually rains back down on the disk.  A number of
arguments have been made to this effect.  \citet{Heckman_et_al_2000}
found the terminal velocity of starburst-driven winds to be
300-900~km~s$^{-1}$, independent of the host galaxy properties, and
invoked simple models of the gravitational potential to suggest that
galactic outflows escape from the potential wells of dwarf galaxies,
but not from more massive hosts.  However these simple scaling
arguments neglect the multi-phase nature of galactic winds, and the
complicated interaction of the wind with the interstellar medium of
the disk and the halo.  Drawing on the numerical simulations of
\citet{MacLow_and_Ferrara_1999} which attempt to include this
complicated gastrophysics, \citet{Ferrara_and_Tolstoy_2000} conclude
that galaxies with gas masses greater than $\sim10^9$~M$_{\sun}$ do
not experience outflows.  In contrast, our results imply that galaxies
with masses as high as $10^{10}$~M$_{\sun}$ still eject some fraction
of their metals into the intergalactic medium.  This result is in line
with the recent observational and theoretical work of
\citet{Strickland_et_al_2003}.  These authors suggest that blowout is
possible from galaxies with masses as high as $10^{11}$~M$_{\sun}$
based on a model where supernovae in disk OB associations work
together to power the outflow.

Of course it is quite likely that even in the absence of winds
chemical evolution does not proceed in the simple manner we have
assumed.  It is worth considering if other factors could explain the
trends we observe.  For example, another means of lowering the
effective yield is the inflow of metal-poor gas.  While inflow may
occur in some galaxies (the Milky Way, for instance) we do not regard
it as a viable explanation for the correlation between effective yield
and baryonic mass.  Metal-poor inflow acts to lower the metallicity
and to raise the gas mass fraction, so the net effect on the effective
yield is modest. For example, consider a galaxy presently undergoing
closed box chemical evolution with a gas fraction of 0.3.  If such a
galaxy experiences an inflow of metal-free gas with a mass equal to
its present gas mass, the metallicity is reduced by 50\%, the gas mass
fraction is increased by 54\%, and the effective yield is decreased by
22\% ($\Delta \log y_{eff}$ = 0.1 dex). Thus, even dramatic inflow
events have only a moderate effect on the effective
yield. \citet{Garnett_2002} pointed out that the low effective yields
of the smallest dwarfs would require them to accrete as much as
80-90\% of their gas at late times without experiencing much star
formation.  We therefore consider it unlikely that metal-poor inflow 
is solely responsible for the low yields we observe.

Another factor that can lower the effective yield is a breakdown of
the instantaneous recycling approximation -- the assumption that gas
is either immediately enriched and returned to the ISM or locked up
forever in low mass stars.  While this assumption obviously does not
hold in detail, the main practical concern is at late times when low
mass stars return their comparatively un-enriched gas to the
interstellar medium.  If this were a chemically significant effect
then we would expect the effective yield to decrease as a function of
mean galaxy age. \citet[][see Fig.~1]{Kauffmann_et_al_2003b}
demonstrated a tight correlation between stellar mass and the
D$_n$(4000) spectral index, which is a good tracer of the mean stellar
age of a population.  The correlation is in the sense that more
massive galaxies appear to be older. Hence we would predict that
$y_{eff}$ should decrease with galaxy mass, just the opposite of what
is seen. While gas recycling may indeed influence the detailed shape
of the $y_{eff} - M_{baryon}$ relation, it is almost certainly not the
main driver of the observed trend.
 
It is our view that the strong positive correlation between
effective yield and baryonic mass is most \emph{naturally} explained
by the increasing potential barrier which the metal-laden wind must
overcome to achieve `blow out'. While the correlation is not
particularly tight ($\pm0.15$~dex), its very existence nevertheless
implies two very interesting things. First, nearly all low mass
galaxies and many high mass galaxies have experienced some sort of
blow-out event.  Since the energy requirements for blow-out are fairly
exacting, it seems likely that these events are associated with
starburst activity.  This chain of reasoning leads us to suggest that
the star formation histories of \emph{most} galaxies may be bursty
rather than continuous.  (However, see \citet{Skillman_2001},
\citet{van_Zee_2001}, and \citet{Hunter_and_Gallagher_1985} for some
opposing viewpoints.)  Unfortunately too little is known about the
extended gaseous halos of normal galaxies and the conditions required
for blow-out to make a more quantitative statement
\citep[see][]{Strickland_et_al_2003}.  The second significant
implication is that blow-out events have an important chemical impact
even when integrated over a galaxy's history of star formation.  This
suggests that blow-out is either a very frequent occurrence or very
catastrophic.  In the coming years numerical and semi-analytical
models should help to address these questions.  In the meantime we
emphasize our empirical findings: galactic winds are ubiquitous and
extremely effective in removing metals from galaxies.  

The quest to understand how the chemical properties of galaxies couple
to their star formation histories has been given added impetus of late
by measurements of the metallicity-luminosity relation of galaxies at
intermediate ($0.3 < z <1.0$) \citep{Kobulnicky_et_al_2003,
Lilly_Carollo_and_Stockton_2003, Maier_et_al_2004} and high ($z > 2$)
redshifts \citep{Kobulnicky_and_Koo_2000, Pettini_et_al_2001}.  Our
results imply that metallicity is not a straightforward metric of
galaxy evolution because metals can escape galactic potential wells.
However, the strong correspondence of metal loss with the size of the
potential has interesting implications if effective yields can be
measured reliably.  The combination of the intermediate and high
redshift data with the correlations we measure at $z\sim0.1$ will
provide an important benchmark for successful models of feedback
and galaxy evolution.

\acknowledgements

The authors wish to thank John Moustakas, Rob Kennicutt, and Roberto
Terlevich for useful discussions, Dave Strickland for helpful
`feedback' on the text, and Larry Bradley for his assistance with our
IDL code. We gratefully acknowledge the referee, Evan Skillman, for
insightful comments that improved the paper.  The data processing
software developed for this project benefited from IDL routines
written by Craig Markwardt and from the IDL Astronomy User's Library
maintained by Wayne Landsman at Goddard Space Flight Center.
C.~A.~T. thanks the Max Planck Institute for Astrophysics and the
Johns Hopkins Center for Astrophysics for their generous financial
support.  She also acknowledges support from NASA grant NAG~58426 and
NSF grant AST-0307386.  J.~B. and S.~C. thank the Alexander von
Humbolt Foundation, the Federal Ministry of Education and Research,
and the Programme for Investment in the Future (ZIP) of the German
Government for their support. J.~B. would also like to acknowledge the
receipt of an ESA post-doctoral fellowship.

Funding for the creation and distribution of the SDSS Archive has been
provided by the Alfred P. Sloan Foundation, the Participating
Institutions, the National Aeronautics and Space Administration, the
National Science Foundation, the U.S. Department of Energy, the
Japanese Monbukagakusho, and the Max Planck Society. The SDSS Web site
is http://www.sdss.org/.
The SDSS is managed by the Astrophysical Research Consortium (ARC) for
the Participating Institutions. The Participating Institutions are The
University of Chicago, Fermilab, the Institute for Advanced Study, the
Japan Participation Group, The Johns Hopkins University, Los Alamos
National Laboratory, the Max-Planck-Institute for Astronomy (MPIA),
the Max-Planck-Institute for Astrophysics (MPA), New Mexico State
University, University of Pittsburgh, Princeton University, the United
States Naval Observatory, and the University of Washington.

\clearpage

\begin{deluxetable}{rrrrrr}
\tabletypesize{\scriptsize}
\tablewidth{0pt}
\tablecaption{The $g$-band Luminosity-Metallicity Relation}
\tablehead{
\colhead{$M_g$} &
\multicolumn{4}{c}{12 + log(O/H)} \\
\cline{2-6} \\
\colhead{} & 
\colhead{$P_{2.5}$} & 
\colhead{$P_{16}$} & 
\colhead{$P_{50}$} & 
\colhead{$P_{84}$} & 
\colhead{$P_{97.5}$} \\ 
}
\startdata
  -22.37  &    8.86  &    8.99 &    9.08  &    9.15  &    9.26  \\
  -21.99  &    8.88  &    9.00 &    9.09  &    9.15  &    9.24  \\
  -21.61  &    8.87  &    8.99 &    9.09  &    9.15  &    9.24  \\
  -21.23  &    8.84  &    8.97 &    9.07  &    9.14  &    9.23  \\
  -20.84  &    8.77  &    8.92 &    9.04  &    9.13  &    9.21  \\
  -20.45  &    8.71  &    8.87 &    9.01  &    9.11  &    9.19  \\
  -20.05  &    8.66  &    8.82 &    8.97  &    9.09  &    9.17  \\
  -19.67  &    8.60  &    8.76 &    8.93  &    9.07  &    9.17  \\
  -19.26  &    8.55  &    8.71 &    8.89  &    9.05  &    9.16  \\
  -18.87  &    8.39  &    8.64 &    8.83  &    9.02  &    9.14  \\
  -18.46  &    8.32  &    8.61 &    8.81  &    9.00  &    9.14  \\
  -18.07  &    8.21  &    8.52 &    8.70  &    8.96  &    9.11  \\
  -17.68  &    8.20  &    8.41 &    8.66  &    8.89  &    9.12  \\
  -17.28  &    8.18  &    8.34 &    8.62  &    8.87  &    9.07  \\
  -16.87  &    8.08  &    8.23 &    8.49  &    8.81  &    9.15  \\
\enddata
 
\tablecomments{The columns labeled $P$ represent the
2.5, 16, 50, 84, and 97.5 percentile of the distribution of
metallicity in bins of 0.4 dex in luminosity. $P_{50}$ is the median.
The $g$-band luminosity is in AB magnitudes and has been corrected for
inclination-dependent extinction but not internal extinction.}
\end{deluxetable}

\begin{deluxetable}{rrrrrr}
\tabletypesize{\scriptsize}
\tablewidth{0pt}
\tablecaption{The $z$-band Luminosity-Metallicity Relation}
\tablehead{
\colhead{$M_{z}^{cor}$} &
\multicolumn{4}{c}{12 + log(O/H)} \\
\cline{2-6} \\
\colhead{} & 
\colhead{$P_{2.5}$} & 
\colhead{$P_{16}$} & 
\colhead{$P_{50}$} & 
\colhead{$P_{84}$} & 
\colhead{$P_{97.5}$} \\ 
}
\startdata
  -23.67  &    9.00  &    9.06 &    9.12  &    9.18  &    9.29  \\
  -23.30  &    8.95  &    9.05 &    9.11  &    9.18  &    9.27  \\
  -22.91  &    8.94  &    9.03 &    9.11  &    9.17  &    9.25  \\
  -22.53  &    8.92  &    9.01 &    9.09  &    9.15  &    9.24  \\
  -22.13  &    8.87  &    8.98 &    9.07  &    9.14  &    9.22  \\
  -21.74  &    8.83  &    8.94 &    9.04  &    9.12  &    9.19  \\
  -21.35  &    8.75  &    8.88 &    9.00  &    9.09  &    9.17  \\
  -20.94  &    8.70  &    8.83 &    8.96  &    9.06  &    9.15  \\
  -20.56  &    8.64  &    8.77 &    8.90  &    9.02  &    9.12  \\
  -20.16  &    8.58  &    8.70 &    8.85  &    8.98  &    9.10  \\
  -19.76  &    8.47  &    8.64 &    8.79  &    8.94  &    9.08  \\
  -19.36  &    8.37  &    8.61 &    8.73  &    8.89  &    9.06  \\
  -18.96  &    8.23  &    8.52 &    8.67  &    8.83  &    9.04  \\
  -18.56  &    8.15  &    8.40 &    8.62  &    8.78  &    9.02  \\
  -18.18  &    8.15  &    8.31 &    8.57  &    8.72  &    8.94  \\
  -17.79  &    8.16  &    8.24 &    8.45  &    8.66  &    8.87  \\
  -17.39  &    8.04  &    8.15 &    8.36  &    8.59  &    8.94  \\
\enddata
 
\tablecomments{The columns labeled $P$ represent the
2.5, 16, 50, 84, and 97.5 percentile of the distribution of
metallicity in bins of 0.4 dex in luminosity. $P_{50}$ is the
median. The $z$-band luminosity is in AB magnitudes and has been
corrected for internal extinction.}
\end{deluxetable}

\begin{deluxetable}{rrrrrr}
\tabletypesize{\scriptsize}
\tablewidth{0pt}
\tablecaption{The Mass--Metallicity Relation}
\tablehead{
\colhead{$\log$(M$_{*}/$M$_{\sun}$)} &
\multicolumn{4}{c}{12 + log(O/H)} \\
\cline{2-6} \\
\colhead{} & 
\colhead{$P_{2.5}$} & 
\colhead{$P_{16}$} & 
\colhead{$P_{50}$} & 
\colhead{$P_{84}$} & 
\colhead{$P_{97.5}$} \\ 
}
\startdata
    8.57  &    8.18  &    8.25 &    8.44  &    8.64  &    8.77  \\
    8.67  &    8.11  &    8.28 &    8.48  &    8.65  &    8.84  \\
    8.76  &    8.13  &    8.32 &    8.57  &    8.70  &    8.88  \\
    8.86  &    8.14  &    8.37 &    8.61  &    8.73  &    8.89  \\
    8.96  &    8.21  &    8.46 &    8.63  &    8.75  &    8.95  \\
    9.06  &    8.26  &    8.56 &    8.66  &    8.82  &    8.97  \\
    9.16  &    8.37  &    8.59 &    8.68  &    8.82  &    8.95  \\
    9.26  &    8.39  &    8.60 &    8.71  &    8.86  &    9.04  \\
    9.36  &    8.46  &    8.63 &    8.74  &    8.88  &    9.03  \\
    9.46  &    8.53  &    8.66 &    8.78  &    8.92  &    9.07  \\
    9.57  &    8.59  &    8.69 &    8.82  &    8.94  &    9.08  \\
    9.66  &    8.60  &    8.72 &    8.84  &    8.96  &    9.09  \\
    9.76  &    8.63  &    8.76 &    8.87  &    8.99  &    9.10  \\
    9.86  &    8.67  &    8.80 &    8.90  &    9.01  &    9.12  \\
    9.96  &    8.71  &    8.83 &    8.94  &    9.05  &    9.14  \\
   10.06  &    8.74  &    8.85 &    8.97  &    9.06  &    9.15  \\
   10.16  &    8.77  &    8.88 &    8.99  &    9.09  &    9.16  \\
   10.26  &    8.80  &    8.92 &    9.01  &    9.10  &    9.17  \\
   10.36  &    8.82  &    8.94 &    9.03  &    9.11  &    9.18  \\
   10.46  &    8.85  &    8.96 &    9.05  &    9.12  &    9.21  \\
   10.56  &    8.87  &    8.98 &    9.07  &    9.14  &    9.21  \\
   10.66  &    8.89  &    9.00 &    9.08  &    9.15  &    9.23  \\
   10.76  &    8.91  &    9.01 &    9.09  &    9.15  &    9.24  \\
   10.86  &    8.93  &    9.02 &    9.10  &    9.16  &    9.25  \\
   10.95  &    8.93  &    9.03 &    9.11  &    9.17  &    9.26  \\
   11.05  &    8.92  &    9.03 &    9.11  &    9.17  &    9.27  \\
   11.15  &    8.94  &    9.04 &    9.12  &    9.18  &    9.29  \\
   11.25  &    8.93  &    9.03 &    9.12  &    9.18  &    9.29  \\
\enddata

\tablecomments{The columns labeled $P$ represent the 2.5, 16, 50, 84,
  and 97.5 percentile of the distribution of metallicity in bins 
  of 0.1 dex in stellar mass. $P_{50}$ is the median.}
\end{deluxetable}

\begin{deluxetable}{rrrrrr}
\tabletypesize{\scriptsize}
\tablewidth{0pt}
\tablecaption{The Baryonic Mass--Effective Yield Relation}
\tablehead{
\colhead{$\log$ (M$_{baryon}$/M$_{\sun}$)} &
\multicolumn{4}{c}{$\log y_{eff}$} \\
\cline{2-6} \\
\colhead{} & 
\colhead{$P_{2.5}$} & 
\colhead{$P_{16}$} & 
\colhead{$P_{50}$} & 
\colhead{$P_{84}$} & 
\colhead{$P_{97.5}$} \\ 
}
\startdata
    8.67  &   -2.84  &   -2.74 &   -2.57  &   -2.34  &   -2.08  \\
    8.76  &   -2.75  &   -2.69 &   -2.52  &   -2.32  &   -2.05  \\
    8.86  &   -2.80  &   -2.68 &   -2.51  &   -2.27  &   -2.03  \\
    8.97  &   -2.74  &   -2.60 &   -2.39  &   -2.18  &   -1.98  \\
    9.06  &   -2.72  &   -2.60 &   -2.40  &   -2.21  &   -1.98  \\
    9.16  &   -2.73  &   -2.53 &   -2.35  &   -2.17  &   -1.98  \\
    9.26  &   -2.69  &   -2.49 &   -2.31  &   -2.13  &   -1.92  \\
    9.36  &   -2.65  &   -2.46 &   -2.30  &   -2.12  &   -1.93  \\
    9.47  &   -2.61  &   -2.43 &   -2.27  &   -2.11  &   -1.95  \\
    9.56  &   -2.56  &   -2.39 &   -2.24  &   -2.08  &   -1.89  \\
    9.66  &   -2.55  &   -2.38 &   -2.24  &   -2.09  &   -1.90  \\
    9.76  &   -2.49  &   -2.35 &   -2.21  &   -2.06  &   -1.90  \\
    9.86  &   -2.50  &   -2.33 &   -2.19  &   -2.04  &   -1.87  \\
    9.96  &   -2.46  &   -2.32 &   -2.18  &   -2.03  &   -1.87  \\
   10.06  &   -2.44  &   -2.30 &   -2.17  &   -2.02  &   -1.85  \\
   10.16  &   -2.44  &   -2.29 &   -2.15  &   -2.01  &   -1.83  \\
   10.26  &   -2.43  &   -2.27 &   -2.13  &   -1.99  &   -1.82  \\
   10.36  &   -2.43  &   -2.26 &   -2.12  &   -1.98  &   -1.80  \\
   10.46  &   -2.41  &   -2.25 &   -2.11  &   -1.96  &   -1.80  \\
   10.56  &   -2.38  &   -2.23 &   -2.09  &   -1.94  &   -1.78  \\
   10.66  &   -2.38  &   -2.22 &   -2.07  &   -1.93  &   -1.77  \\
   10.76  &   -2.36  &   -2.21 &   -2.07  &   -1.92  &   -1.78  \\
   10.86  &   -2.36  &   -2.21 &   -2.06  &   -1.91  &   -1.77  \\
   10.95  &   -2.36  &   -2.20 &   -2.05  &   -1.90  &   -1.75  \\
   11.05  &   -2.38  &   -2.20 &   -2.05  &   -1.90  &   -1.74  \\
   11.15  &   -2.36  &   -2.19 &   -2.04  &   -1.88  &   -1.70  \\
   11.25  &   -2.38  &   -2.20 &   -2.03  &   -1.86  &   -1.73  \\
   11.35  &   -2.30  &   -2.18 &   -2.01  &   -1.82  &   -1.68  \\
\enddata

\tablecomments{The columns labeled $P$ represent the 2.5, 16, 50, 84,
  and 97.5 percentile of the distribution of the effective yield in bins 
  of 0.1 dex in baryonic mass. $P_{50}$ is the median.}
\end{deluxetable}

\end{document}